\documentclass[12pt]{article}

\usepackage{latexsym}
\usepackage{times}

\input{psfig.sty}

\begin{document}

\newcommand{\mycaption}[2]{\caption[#1]{{\bf #1} #2}}

\newcommand{\comment}[1]{#1}

\newcommand{\fsat}{F_{\mbox{sat}}}
\newcommand{\fver}{F_{\mbox{ver}}}
\newcommand{\fsver}{F^*_{\mbox{ver}}}
\newcommand{\compare}[2]{#1\!=\!#2} 
\newcommand{\transset}{{\it Trans}}
\newcommand{\transfunct}{{\it T}}
\newcommand{\varset}{{\cal V}}
\newcommand{\tvarset}{{\cal E}}
\newcommand{\atvarset}{\tvarset^{+}}
\newcommand{\rtvarset}{\hat{\tvarset}}
\newcommand{\artvarset}{\hat{\tvarset}^{+}}
\newcommand{\transclauseset}{{\it C}_{\mbox{trans}}}
\newcommand{\ftrans}{F_{\mbox{trans}}}
\newcommand{\formclauseset}{{\it C}_{\mbox{sat}}}
\newcommand{\asst}{\chi}
\newcommand{\aeval}[2]{\left[ #1 \right]_{#2}}
\newcommand{\meshfunct}{{\it MT}}
\newcommand{\meshvars}{\tvarset_{n \times n}}
\newcommand{\encfunct}{{\it encf}}
\newcommand{\mesh}[1]{M_{#1}}
\newcommand{\meshn}{\mesh{n}}
\newcommand{\dualmesh}[1]{M^D_{#1}}
\newcommand{\dualmeshn}{\dualmesh{n}}

\comment{
\newtheorem{lemma}{Lemma}
\newtheorem{theorem}{Theorem}
\newtheorem{proposition}{Proposition}
\newtheorem{corollary}{Corollary}
\newenvironment{proof}{{\it Proof:}}{\qed}
\newcommand{\qed}{\mbox{} $\Box$}
}

\newcommand{\clabel}[1]{\makebox(0,0){#1}}
\newcommand{\rlabel}[1]{\makebox(0,0)[r]{#1}}
\newcommand{\llabel}[1]{\makebox(0,0)[l]{#1}}
\newcommand{\blabel}[1]{\makebox(0,0)[b]{#1}}
\newcommand{\tlabel}[1]{\makebox(0,0)[t]{#1}}
\newcommand{\vertex}{\makebox(0,0){$\bullet$}}

\newcommand{\dlxc}{\mbox{1$\times$DLX-C}}
\newcommand{\dlxct}{\mbox{1$\times$DLX-C-t}}
\newcommand{\dlxca}{\mbox{2$\times$DLX-CA}}
\newcommand{\dlxcat}{\mbox{2$\times$DLX-CA-t}}
\newcommand{\dlxcc}{\mbox{2$\times$DLX-CC}}
\newcommand{\dlxcct}{\mbox{2$\times$DLX-CC-t}}

\title{Boolean Satisfiability \\ with Transitivity Constraints\thanks{Supported, in part, by the
Semiconductor Research Corporation under contract 00-DC-684.}}

\author{
Randal E. Bryant\\
Miroslav N. Velev\\
Carnegie Mellon University}

\maketitle

\begin{abstract}
We consider a variant of the Boolean satisfiability problem where a
subset $\tvarset$ of the propositional variables appearing in formula
$\fsat$ encode a symmetric, transitive, binary relation over $N$
elements.  Each of these {\em relational} variables, 
$e_{i,j}$, for $1 \leq i < j \leq N$, expresses whether or not the
relation holds between elements $i$ and $j$.  The task is to either
find a satisfying assignment to $\fsat$ that also satisfies all
transitivity constraints over the relational variables (e.g., $e_{1,2}
\land e_{2,3} \Rightarrow e_{1,3}$), or to prove that no such
assignment exists.  Solving this satisfiability problem is the final
and most difficult step in our decision procedure for a logic of
equality with uninterpreted functions.  This procedure forms the core
of our tool for verifying pipelined microprocessors.

To use a conventional Boolean satisfiability checker, we augment the
set of clauses expressing $\fsat$ with clauses expressing the
transitivity constraints.  We consider methods to reduce the number of
such clauses based on the sparse structure of the relational
variables.

To use Ordered Binary Decision Diagrams (OBDDs), we show that for some
sets $\tvarset$, the OBDD representation of the transitivity
constraints has exponential size for all possible variable orderings.
By considering only those relational variables that occur in the
OBDD representation of $\fsat$, our experiments show that we can
readily construct an OBDD representation of the relevant transitivity
constraints and thus solve the constrained satisfiability problem.

\end{abstract}

\section{Introduction}

Consider the following variant of the Boolean satisfiability problem.
We are given a Boolean formula $\fsat$ over a set of variables
$\varset$.  A subset $\tvarset \subseteq \varset$ symbolically encodes
a binary relation over $N$ elements
that is reflexive, symmetric, and transitive.  Each of
these {\em relational} variables, $e_{i,j}$, where $1
\leq i < j \leq N$, expresses whether or not the relation holds
between elements $i$ and $j$.  Typically, $\tvarset$ will be
``sparse,'' containing much fewer than the $N(N-1)/2$ possible
variables.  Note that when $e_{i,j} \not \in \tvarset$ for some value
of $i$ and of $j$, this does not imply that the relation does not hold
between elements $i$ and $j$.  It simply indicates that $\fsat$ does
not directly depend on the relation between elements $i$ and $j$.

A {\em transitivity
constraint} is a formula of the form
\begin{eqnarray}
e_{[i_1,i_2]}
\land e_{[i_2,i_3]} \land \cdots \land e_{[i_{k-1},i_{k}]} &
\Rightarrow &
e_{[i_1,i_k]}
\label{transitivity-equation}
\end{eqnarray}
where $e_{[i,j]}$ equals $e_{i,j}$ when $i < j$
and equals $e_{j,i}$ when $i > j$.  Let $\transset(\tvarset)$ denote
the set of all transitivity constraints that can be formed from the
relational variables.  Our task is to find an assignment $\asst \colon
\varset \rightarrow \{0,1\}$ that satisfies $\fsat$, as well as every
constraint in $\transset(\tvarset)$.  Goel, {\em et
al.}~\cite{goel-cav98} have shown this problem is NP-hard, even when
$\fsat$ is given as an Ordered Binary Decision Diagram (OBDD)
\cite{bryant-ieeetc86}.  Normally, Boolean satisfiability is trivial
given an OBDD representation of a formula.

We are motivated to solve this problem as part of a tool for verifying
pipelined microprocessors \cite{velev-charme99}.  Our tool abstracts
the operation of the datapath as a set of uninterpreted functions and
uninterpreted predicates operating on symbolic data.  We prove that a
pipelined processor has behavior matching that of an unpipelined
reference model using the symbolic flushing technique developed by
Burch and Dill \cite{burch-cav94}.  The major computational task is to
decide the validity of a formula $\fver$ in a logic of equality with
uninterpreted functions \cite{bryant-cav99,bryant-tr99}.  Our decision
procedure transforms $\fver$ first by replacing all function
application terms with terms over a set of domain variables $\{v_i | 1
\leq i \leq N\}$.  Similarly, all predicate applications are replaced
by formulas over a set of newly-generated propositional variables.
The result is a formula $\fsver$ containing equations of the form
$\compare{v_i}{v_j}$, where $1 \leq i < j \leq N$.  Each of these
equations is then encoded by introducing a relational variable
$e_{i,j}$, similar to the method proposed by Goel, {\em et
al.}~\cite{goel-cav98}.  The result of the translation is a
propositional formula ${\it encf}(\fsver)$ expressing the verification
condition over both the relational variables and the
propositional variables appearing in $\fsver$.  Let $\fsat$ denote
$\neg {\it encf}(\fsver)$, the complement of the formula expressing
the translated verification condition.  To capture the transitivity of
equality, e.g., that $\compare{v_i}{v_j} \land \compare{v_j}{v_k}
\Rightarrow \compare{v_i}{v_k}$, we have transitivity constraints of
the form $e_{[i,j]} \land e_{[j,k]} \Rightarrow e_{[i,k]}$.  Finding a
satisfying assignment to $\fsat$ that also satisfies the transitivity
constraints will give us a counterexample to the original verification
condition $\fver$.  On the other hand, if we can prove that there are
no such assignments, then we have proved that $\fver$ is universally
valid.

We consider three methods to generate a Boolean formula $\ftrans$ that
encodes the transitivity constraints. The {\em direct}
method enumerates the set of {\em chord-free} cycles in the undirected
graph having an edge $(i,j)$ for each relational variable $e_{i,j} \in
\tvarset$.  This method avoids introducing additional relational variables
but can lead to a formula of exponential size.  The {\em dense} method
uses relational variables $e_{i,j}$ for all possible values of
$i$ and $j$ such that $1 \leq i < j \leq N$.  We can then axiomatize
transitivity by forming constraints of the form $e_{[i,j]} \land
e_{[j,k]} \Rightarrow e_{[i,k]}$ for all distinct values of $i$, $j$,
and $k$.  This will yield a formula that is cubic in $N$.  The {\em
sparse} method augments $\tvarset$ with additional relational
variables to form a set of variables $\atvarset$, such that the
resulting graph is {\em chordal} \cite{rose-jmaa70}.  We then only
require transitivity constraints of the form $e_{[i,j]} \land
e_{[j,k]} \Rightarrow e_{[i,k]}$ such that $e_{[i,j]}, e_{[j,k]},
e_{[i,k]} \in \atvarset$.  The sparse method is guaranteed to generate
a smaller formula than the dense method.

To use a conventional Boolean Satisfiability (SAT) procedure to
solve our constrained satisfiability problem, we run the checker
over a set of clauses encoding both $\fsat$ and $\ftrans$.
The latest version of the {\sc fgrasp} SAT checker \cite{marques-pcai99} was able to complete
all of our benchmarks, although the run times increase
significantly when transitivity constraints are enforced.

When using Ordered Binary Decision Diagrams to evaluate satisfiability,
we could generate OBDD representations of $\fsat$ and $\ftrans$ and use the
{\sc apply} algorithm to compute an OBDD representation of their
conjunction.  From this OBDD, finding satisfying solutions would be
trivial.  We show that this approach will not be feasible in general,
because the OBDD representation of $\ftrans$ can be intractable.  That
is, for some sets of relational variables, the OBDD representation of
the transitivity constraint formula $\ftrans$ will be of exponential
size regardless of the variable ordering.  The NP-completeness result
of Goel, {\em et al.}~shows that the OBDD representation of $\ftrans$
may be of exponential size using the ordering previously selected for
representing $\fsat$ as an OBDD.  This leaves open the possibility
that there could be some other variable ordering that would yield
efficient OBDD representations of both $\fsat$ and $\ftrans$.  Our result
shows that transitivity constraints can be intrinsically intractable
to represent with OBDDs, independent of the structure of $\fsat$.

We present experimental results on the complexity of constructing
OBDDs for the transitivity constraints that arise in actual
microprocessor verification.  Our results show that the OBDDs can
indeed be quite large.  We consider two techniques to avoid
constructing the OBDD representation of all transitivity constraints.
The first of these, proposed by Goel, {\em et al.}~\cite{goel-cav98},
generates implicants (cubes) of $\fsat$ and rejects those that violate
the transitivity constraints.  Although this method suffices for small
benchmarks, we find that the number of implicants generated for our
larger benchmarks grows unacceptably large.  The second method
determines which relational variables actually occur in the OBDD
representation of $\fsat$.  We can then apply one of our three
techniques for encoding the transitivity constraints in order to
generate a Boolean formula for the transitivity constraints over this
reduced set of relational variables.  The OBDD representation of this
formula is generally tractable, even for the larger benchmarks.

\section{Benchmarks}

\begin{table}
\begin{center}
\begin{tabular}{|l|l|rrr|}
\hline
\multicolumn{2}{|c|}{Circuit} & \multicolumn{1}{|c}{Domain}  & 
\multicolumn{1}{c}{Propositional} & 
\multicolumn{1}{c|}{Equations} \\
\multicolumn{2}{|c}{} & \multicolumn{1}{|c}{Variables} & 
\multicolumn{1}{c}{Variables} & \\
\hline
\multicolumn{2}{|l|}{\dlxc} & 13 & 42 & 27 \\
\multicolumn{2}{|l|}{\dlxct} & 13 & 42 & 37 \\
\hline
\multicolumn{2}{|l|}{\dlxca} & 25 & 58 & 118 \\
\multicolumn{2}{|l|}{\dlxcat} & 25 & 58 & 137 \\
\hline
\multicolumn{2}{|l|}{\dlxcc} & 25 & 70 & 124 \\
\multicolumn{2}{|l|}{\dlxcct} & 25 & 70 & 143 \\
\hline
Buggy & min. & 22 & 56 & 89\\
\dlxcc{} & avg. & 25 & 69 & 124\\
 & max. & 25 & 77 & 132\\
\hline
\end{tabular}
\end{center}
\mycaption{Microprocessor Verification Benchmarks.}{
Benchmarks with suffix ``t'' were modified to require 
enforcing transitivity.}
\label{benchmark-table}
\end{table}

Our benchmarks \cite{velev-charme99} are based on applying our
verifier to a set of high-level microprocessor designs.  Each is based
on the DLX RISC processor described by Hennessy and Patterson
\cite{hennessy-96}:
\begin{description}
\item[\dlxc{}:] is a single-issue, five-stage pipeline capable of
 fetching up to one new instruction every clock cycle.  It implements
 six instruction types:
 register-register, register-immediate, load,
 store, branch, and jump. The pipeline stages are: Fetch, Decode,
 Execute, Memory, and Write-Back.  An interlock causes the instruction
 following a load to stall one cycle if it requires the loaded result.
 Branches and jumps are predicted as not-taken, with up to 3
 instructions squashed when there is a misprediction.
This example is
comparable to the DLX example first verified by Burch and Dill
 \cite{burch-cav94}.

\item[\dlxca{}:] has a complete first pipeline, capable of executing
the six instruction types, and a second pipeline capable of executing
arithmetic instructions.
\comment{  Between 0 and
2 new instructions are issued on each cycle, depending on their types
and source registers, as well as the types and destination registers
of the preceding instructions.} This example is comparable to one
verified by Burch \cite{burch-dac96}.

\item[\dlxcc{}:] has two complete pipelines, i.e., each can execute any of the six instruction types.
\comment{There are four load
interlocks---between a load in Execute in either pipeline and an
instruction in Decode in either pipeline.  On each cycle, between 0
and 2 instructions can be issued.}
\end{description}
In all of these examples, the domain variables $v_i$, with $1 \leq i
 \leq N$, in $\fsver$ encode register identifiers.  As described in
 \cite{bryant-cav99,bryant-tr99}, we can encode the symbolic terms
 representing program data and addresses as distinct values, avoiding
 the need to have equations among these variables.  Equations
 arise in modeling the read and write
 operations of the register file, the bypass logic implementing
 data forwarding, the load interlocks, and the pipeline issue
 logic.

Our original processor benchmarks can be verified without 
enforcing any transitivity constraints.
The unconstrained formula $\fsat$ is
unsatisfiable in every case.
We are nonetheless
motivated to study the problem of constrained satisfiability for two
reasons.  First, other processor designs might rely on transitivity,
e.g., due to more sophisticated issue logic.  Second, to aid designers
in debugging their pipelines, it is essential that we generate
counterexamples that satisfy all transitivity constraints.  Otherwise
the designer will be unable to determine whether the counterexample
represents a true bug or a weakness of our verifier.

To create more challenging benchmarks, we generated variants of the
circuits that require enforcing transitivity in the verification.  For
example, the normal forwarding logic in the Execute stage of
\dlxc{} 
must determine whether to forward the result
from the Memory stage instruction as either one or both operand(s) for the
Execute stage instruction.  It does this by
comparing the two source registers $\mbox{\tt ESrc1}$
and $\mbox{\tt ESrc2}$ of the instruction in the Execute stage to the
destination register $\mbox{\tt MDest}$ of the instruction in the
memory stage.   In the modified circuit, we changed the
bypass condition $\compare{\mbox{\tt ESrc1}}{\mbox{\tt MDest}}$ to be
$\compare{\mbox{\tt ESrc1}}{\mbox{\tt MDest}} \lor (\compare{\mbox{\tt
ESrc1}}{\mbox{\tt ESrc2}} \land \compare{\mbox{\tt ESrc2}}{\mbox{\tt
MDest}})$.  Given transitivity, these two expressions are equivalent.
For each pipeline, we introduced four such modifications to the
forwarding logic, with different combinations of source and
destination registers.  These modified circuits are named
\dlxct{}, \dlxcat{}, and \dlxcct{}.

To study the problem of counterexample generation for buggy circuits,
we generated 105 variants of \dlxcc{}, each containing a small
modification to the control logic.  Of these, 5 were found to be
functionally correct, e.g., because the modification caused the
processor to stall unnecessarily, yielding a total of 100 benchmark
circuits for counterexample generation.

Table \ref{benchmark-table} gives some statistics for the
 benchmarks.  The number of domain variables $N$ ranges between 13 and 25,
 while the number of equations ranges between 27 and 143.  The
 verification condition formulas $\fsver$ also contain between 42 and
 77 propositional variables expressing the operation of the control
 logic.  These variables plus the relational variables 
 comprise the set of variables $\varset$ in the propositional formula
 $\fsat$.  The circuits with modifications that require
 enforcing transitivity yield formulas containing up to 19
 additional equations.  The final three lines summarize the complexity
 of the 100 buggy variants of \dlxcc{}.  We apply a number of
 simplifications during the generation of formula $\fsat$, and hence
 small changes in the circuit can yield significant variations in the
 formula complexity.

\section{Graph Formulation}
Our definition of $\transset(\tvarset)$ (Equation
 \ref{transitivity-equation}) places no restrictions on the
 length or form of the transitivity constraints, and hence there can
 be an infinite number.  We show that we can construct a graph
 representation of the relational variables and identify a reduced set
 of transitivity constraints that, when satisfied, guarantees that all
 possible transitivity constraints are satisfied.  By introducing more
 relational variables, we can alter this graph structure, further
 reducing the number of transitivity constraints that must be
 considered.

For variable set $\tvarset$, define the undirected graph
$G(\tvarset)$ as containing a vertex $i$ for $1 \leq i \leq N$, and an
edge $(i,j)$ for each variable $e_{i,j} \in \tvarset$.  For an
assignment $\asst$ of Boolean values to the relational variables,
define the labeled graph $G(\tvarset,\asst)$ to be the graph
$G(\tvarset)$ with
each edge $(i,j)$ labeled as a {\em 1-edge} when $\asst(e_{i,j}) =
1$, and as a {\em 0-edge} when $\asst(e_{i,j}) = 0$.

A {\em path} is a sequence of vertices $[i_1, i_2, \ldots, i_k]$
having edges between successive elements.  That is, each element $i_p$
of the sequence ($1 \leq p \leq k$) denotes a vertex: $1 \leq i_p \leq
N$, while each successive pair of elements $i_p$ and $i_{p+1}$ ($1
\leq p < k$) forms an edge $(i_p, i_{p+1})$ We consider each edge
$(i_p, i_{p+1})$ for $1 \leq p < k$ to also be part of the path.  A
{\em cycle} is a path of the form $[i_1, i_2, \ldots, i_k, i_1]$.

\begin{proposition}
An assignment $\asst$ to the variables in $\tvarset$ violates transitivity if and only if
some cycle in $G(\tvarset,\asst)$ contains exactly one 0-edge.
\label{cycle-proposition}
\end{proposition}

\comment{
\begin{proof}
{\em If}. Suppose there is such a cycle.  Letting $i_1$ be the vertex
at one end of the 0-edge, we can trace around the cycle, giving a
sequence of vertices $[i_1, i_2, \ldots, i_k]$, where $i_k$ is the
vertex at the other end of the 0-edge.  The assignment has
$\asst(e_{[i_j,i_{j+1}]}) = 1$ for $1 \leq j < k$, and
$\asst(e_{[i_1,i_k]} = 0)$, and hence it violates Equation
\ref{transitivity-equation}.

{\em Only If}. Suppose the assignment violates a transitivity
constraint given by Equation \ref{transitivity-equation}.  Then, we
construct a cycle $[i_1, i_2, \ldots, i_k, i_1]$ of vertices such that
only edge $(i_k,i_1)$ is a 0-edge.
\end{proof}
}

A path $[i_1, i_2, \ldots, i_k]$ is said to be {\em acyclic} when $i_p
\not = i_q$ for all $1 \leq p < q \leq k$.  A cycle $[i_1, i_2,
\ldots, i_k, i_1]$ is said to be {\em simple} when its prefix $[i_1,
i_2, \ldots, i_k]$ is acyclic.

\begin{proposition}
An assignment $\asst$ to the variables in $\tvarset$ violates transitivity if and
only if some {\em simple} cycle in $G(\tvarset,\asst)$ contains exactly one 0-edge.
\label{simple-cycle-proposition}
\end{proposition}

\comment{
\begin{proof}
The ``if'' portion of this proof is covered by Proposition
\ref{cycle-proposition}.  The ``only if'' portion is proved by
induction on the number of variables in the antecedent of the
transitivity constraint (Equation \ref{transitivity-equation}.)  That
is, assume a transitivity constraint containing $k$ variables in the
antecedent is violated and that all other violated constraints have at
least $k$ variables in their antecedents.  If there are no values $p$
and $q$ such that $1 \leq p < q \leq k$ and $i_p = i_q$, then the
cycle $[i_1, i_2, \ldots i_k, i_1]$ is simple.  If such values $p$ and
$q$ exist, then we can form a transitivity constraint:
\begin{eqnarray*}
e_{[i_1,i_2]} \land \cdots \land e_{[i_{p-1},i_{p}]} \land e_{[i_{q},i_{q+1}]}
\land \cdots  \land e_{[i_{k-1},i_{k}]} 
& \Rightarrow & e_{[i_{1},i_{k}]} 
\end{eqnarray*}
This transitivity constraint contains fewer than $k$ variables in the
antecedent, but it is also violated.  This contradicts our assumption
that there is no violated transitivity constraint with fewer than $k$ variables in the
antecedent.
\end{proof}
}

Define a {\em chord} of a simple cycle to be an edge that connects two vertices that are not adjacent
in the cycle.  More precisely, for a simple cycle $[i_1, i_2, \ldots, i_k, i_1]$,
a chord is an edge $(i_p, i_q)$ in $G(\tvarset)$ such that $1 \leq p < q \leq
k$, with $p+1 < q$, and either $p \not= 1$ or $q \not = k$.
A cycle is said to be {\em chord-free}
if it is simple and has no chords.

\begin{proposition}
An assignment $\asst$ to the variables in $\tvarset$ violates transitivity if
and only if some {\em chord-free} cycle in $G(\tvarset,\asst)$ contains
exactly one 0-edge.
\label{chord-free-cycle-proposition}
\end{proposition}

\begin{figure}
\centerline{\psfig{figure=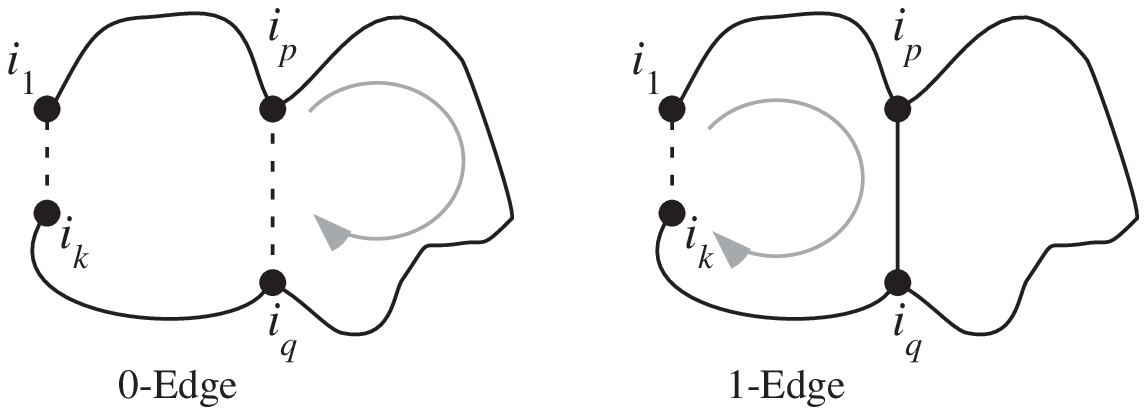}}
\mycaption{Case Analysis for Proposition
\protect{\ref{chord-free-cycle-proposition}}.}{0-Edges are shown as dashed lines.  When a cycle representing a
transitivity violation contains a chord, we can find a smaller cycle that
also represents a transitivity violation.}
\label{chord-split-figure}
\end{figure}

\comment{
\begin{proof}
The ``if'' portion of this proof is covered by Proposition
\ref{cycle-proposition}.  The ``only if'' portion is proved by
induction on the number of variables in the antecedent of the
transitivity constraint (Equation \ref{transitivity-equation}.)
Assume a transitivity constraint with $k$ variables is violated, and
that no transitivity constraint with fewer variables in the antecedent
is violated.  If there are no values of $p$ and $q$ such that there is
a variable $e_{[i_p,i_q]} \in \tvarset$ with $p+1 < q$ and either $p
\not = 1$ or $q \not = k$, then the corresponding cycle is chord-free.
If such values of $p$ and $q$ exist, then consider the two cases
illustrated in Figure \ref{chord-split-figure}, where 0-edges are
shown as dashed lines, 1-edges are shown as solid lines, and the wavy
lines represent sequences of 1-edges.  Case 1: Edge $(i_p,i_q)$ is a
0-edge (shown on the left).  Then the transitivity constraint:
\begin{eqnarray*}
e_{[i_p,i_{p+1}]} \land \cdots \land e_{[i_{q-1},i_{q}]}
& \Rightarrow & e_{[i_{p},i_{q}]} 
\end{eqnarray*}
is violated and has fewer than $k$ variables in its antecedent.
Case 2: Edge $(i_p,i_q)$ is a 1-edge (shown on the right).  Then the transitivity constraint:
\begin{eqnarray*}
e_{[i_1,i_2]} \land \cdots \land e_{[i_{p-1},i_{p}]} \land e_{[i_p,i_q]} \land
e_{[i_{q},i_{q+1}]}
\land \cdots \land
e_{[i_{k-1},i_{k}]}
& \Rightarrow & e_{[i_{1},i_{k}]} 
\end{eqnarray*}
is violated and has fewer than $k$ variables.  Both cases contradict
our assumption that there is no violated transitivity constraint with
fewer than $k$ variables in the antecedent.
\end{proof}
}

Each length $k$ cycle $[i_1, i_2, \ldots,
i_k, i_1]$ yields $k$ constraints,
given by the following clauses.  Each clause is derived by expressing Equation \ref{transitivity-equation} as a disjunction.
\begin{equation}
\begin{array}{l}
\neg e_{[i_{1},i_{2}]} \lor \cdots  \lor \neg e_{[i_{k-1},i_{k}]}
\lor  e_{[i_{k},i_{1}]}  \\
\neg e_{[i_{2},i_{3}]} \lor \cdots \lor \neg e_{[i_{k-1},i_{k}]}
\lor \neg e_{[i_{k},i_{1}]} \lor
  e_{[i_{1},i_{2}]}   \\
\;\;\;\ldots\\
 \neg e_{[i_{k},i_{1}]} \lor
\neg e_{[i_{1},i_{2}]} \lor \cdots \lor \neg e_{[i_{k-2},i_{k-1}]} \lor
  e_{[i_{k-1},i_{k}]} \\
\end{array}
\label{constraint-equation}
\end{equation}
For a set of relational variables $\tvarset$, we define
$\ftrans(\tvarset)$ to be the conjunction of all transitivity
constraints for all chord-free cycles
in the graph $G(\tvarset)$.

\begin{theorem}
An assignment to the relational variables $\tvarset$ will satisfy all of the
transitivity constraints given by Equation \ref{transitivity-equation}
 if and only if it satisfies
$\ftrans(\tvarset)$.
\label{ftrans-theorem}
\end{theorem}

This theorem follows directly from
Proposition \ref{chord-free-cycle-proposition}
and the encoding given by Equation \ref{constraint-equation}.

\subsection{Enumerating Chord-Free Cycles}

To enumerate the chord-free cycles of a graph, we exploit the following
properties.  An acyclic path $[i_1, i_2, \ldots, i_k]$ is said to have
a chord when there is an edge $(i_p, i_q)$ in $G(\tvarset)$ such that
$1 \leq p < q \leq k$ with $p+1 < q$, and either $p \not = 1$ or $q
\not = k$.  We classify a chord-free path as {\em terminal} when
$(i_k, i_1)$ is in $G(\tvarset)$, and as {\em extensible} otherwise.

\begin{proposition}
A path $[i_1, i_2, \ldots, i_k]$ is chord-free and terminal if and only if
the cycle $[i_1, i_2, \ldots, i_k, i_1]$ is chord-free.
\end{proposition}
This follows by noting that the conditions imposed on a chord-free path are
identical to those for a chord-free cycle, except that the latter includes
a closing edge $(i_k, i_1)$.

A {\em proper prefix} of path $[i_1, i_2, \ldots, i_k]$ is a path
$[i_1, i_2, \ldots, i_j]$ such that $1 \leq j < k$.
\begin{proposition}
Every proper prefix  of a chord-free
path  is chord-free and extensible.
\end{proposition}

Clearly, any prefix of a chord-free path is also chord-free.  If some
prefix $[i_1, i_2, \ldots, i_j]$ with $j < k$ 
were terminal, then any attempt to add the edge $(i_j, i_{j+1})$
would yield either a simple cycle (when $i_{j+1} = i_1$), some other
cycle (when $i_{j+1} = i_p$ for some $1 < p < j$), or a path having
$(i_1, i_j)$ as a chord.

Given these properties, we can enumerate the set of all chord-free
paths by breadth first expansion.  As we enumerate these paths, we
also generate $C$, the set of all chord-free cycles.  Define $P_k$ to
be the set of all extensible, chord-free paths having $k$ vertices,
for $1 \leq k \leq N$.

Initially we have $P_1 = \{[i] | 1 \leq i \leq n\}$, and $C =
\emptyset$.  
Given set $P_{k}$, we generate set $P_{k+1}$ and
add some cycles of length $k+1$ to $C$.  For each
path $[i_1, i_2, \ldots, i_{k}] \in P_{k}$, we consider the path
$[i_1, i_2, \ldots, i_{k}, i_{k+1}]$ for each edge $(i_{k}, i_{k+1})$
in $G(\tvarset)$.  When $i_{k+1} = i_p$ for some $1 \leq p < k$, we
classify the path as cyclic.  When there is an edge $(i_{k+1}, i_p)$
in $G(\tvarset)$ for some $1 < p < k$, we classify the path as having
a chord.  When there is an edge $(i_{k+1}, i_1)$ in $G(\tvarset)$, we
add the cycle $[i_1, i_2, \ldots, i_{k}, i_{k+1}, i_1]$
to $C$.  Otherwise, we add the path to $P_{k+1}$.

After generating all of these paths, we can use the set $C$
to generate the set of all chord-free cycles.  For each terminal,
chord-free cycle having  $k$ vertices, there will be $2k$ members of
$C$---each of the $k$ edges of the cycle can serve as the
closing edge, and a cycle can traverse the closing edge in either
direction.  To generate the set of clauses given by Equation
\ref{constraint-equation}, we simply need to choose one element of
$C$ for each closing edge, e.g., by considering only cycles
$[i_1, \ldots, i_k, i_1]$ for which $i_1 < i_k$.

\begin{figure}[t]
\setlength{\unitlength}{.2in}
\begin{center}
\newcommand{\fdiamond}[1]{
  \put(-2,0){\vertex}
  \put(0,2){\vertex}
  \put(0,-2){\vertex}
  \put(2,0){\vertex}
  \put(-2,0){\line(1,1){2}}
  \put(-2,0){\line(1,-1){2}}
  \put(0,2){\line(1,-1){2}}
  \put(0,-2){\line(1,1){2}}
  \put(0,0){\clabel{$F_{#1}$}}
}
\begin{picture}(17,6)
\put(3,4){\fdiamond{1}}
\put(7,4){\fdiamond{2}}
\put(10.5,4){\clabel{$\cdots$}}
\put(14,4){\fdiamond{n}}
\put(1,2){\oval(2,4)[l]}
\put(1,0){\line(1,0){15}}
\put(16,2){\oval(2,4)[r]}
\end{picture}
\end{center}
\mycaption{Class of Graphs with Many Chord-Free Cycles.}{For a graph
with $n$ diamond-shaped faces, there are $2^n+n$ chord-free cycles.}
\label{diamond-figure}
\end{figure}

As Figure \ref{diamond-figure} indicates, there can be an exponential
number of chord-free cycles in a graph.  In particular, this figure
illustrates a family of graphs with $3n+1$ vertices.  Consider the
cycles passing through the $n$ diamond-shaped faces as well as the
edge along the bottom.  For each diamond-shaped face $F_i$, a cycle
can pass through either the upper vertex or the lower vertex.  Thus
there are $2^n$ such cycles.  In addition, the edges forming the
perimeter of each face $F_i$ create a chord-free cycle, giving a
total of $2^n + n$ chord-free cycles.

\begin{table}
\begin{center}
{\small
\begin{tabular}{|l|l|rrr|rrr|rrr|}
\hline
\multicolumn{2}{|c|}{Circuit}  
 & \multicolumn{3}{|c|}{Direct} &
\multicolumn{3}{|c|}{Dense} & \multicolumn{3}{|c|}{Sparse} \\
\multicolumn{2}{|c|}{} & \multicolumn{1}{|c}{Edges} & 
\multicolumn{1}{c}{Cycles} & 
\multicolumn{1}{c|}{Clauses} &
\multicolumn{1}{|c}{Edges} & 
\multicolumn{1}{c}{Cycles} & 
\multicolumn{1}{c|}{Clauses} & 
\multicolumn{1}{|c}{Edges} & 
\multicolumn{1}{c}{Cycles} & 
\multicolumn{1}{c|}{Clauses} \\
\hline
\multicolumn{2}{|l|}{\dlxc}  & 27 & 90 & 360 & 78 & 286 & 858 & 33 & 40 & 120  \\
\multicolumn{2}{|l|}{\dlxct}  & 37 & 95 & 348 & 78 & 286 & 858& 42 & 68 & 204  \\
\hline
\multicolumn{2}{|l|}{\dlxca}  & 118 & 2,393 & 9,572 & 300 & 2,300 & 6,900 & 172 & 697 & 2,091 \\
\multicolumn{2}{|l|}{\dlxcat} & 137 & 1,974 & 7,944 & 300 & 2,300 & 6,900 & 178 & 695 & 2,085 \\
\hline
\multicolumn{2}{|l|}{\dlxcc}  & 124 & 2,567 & 10,268 & 300 & 2,300 & 6,900 & 182 & 746 & 2,238  \\
\multicolumn{2}{|l|}{\dlxcct} & 143 & 2,136 &  8,364 & 300 & 2,300 & 6,900 & 193 & 858 & 2,574  \\
\hline
Full  & min. & 89 & 1,446 & 6,360 & 231 & 1,540 & 4,620 & 132 & 430 & 1,290 \\
Buggy & avg. & 124 & 2,562 & 10,270 & 300 & 2,300 & 6,900 & 182 & 750 & 2,244  \\
\dlxcc & max.  & 132 & 3,216 & 12,864 & 299 & 2,292 & 6,877 & 196 & 885 & 2,655 \\
\hline
\multicolumn{2}{|l|}{$M_4$}  & 24 & 24 & 192 & 120 & 560 & 1,680 & 42 & 44 & 132 \\
\multicolumn{2}{|l|}{$M_5$}  & 40 & 229 & 3,056 & 300 & 2,300 & 6,900 & 77 & 98 & 294  \\
\multicolumn{2}{|l|}{$M_6$}  & 60 & 3,436 & 61,528 & 630 & 7,140 & 21,420 & 131 & 208 & 624 \\
\multicolumn{2}{|l|}{$M_7$}  & 84 & 65,772 & 1,472,184 & 1,176 & 18,424 & 55,272 & 206 & 408 & 1,224  \\
\multicolumn{2}{|l|}{$M_8$}  & 112 & 1,743,247 & 48,559,844 & 2,016 & 41,664 & 124,992 & 294 & 662 & 1,986  \\
\hline
\end{tabular}
} 
\end{center}
\mycaption{Cycles in Original and Augmented Benchmark Graphs.}{Results are given for the three different methods
of encoding transitivity constraints.}
\label{chord-free-table}
\end{table}

The columns labeled ``Direct'' in Table \ref{chord-free-table} show
results for enumerating the chord-free cycles for our benchmarks.  For
each correct microprocessor, we have two graphs: one for which
transitivity constraints played no role in the verification, and one
(indicated with a ``t'' at the end of the name) modified to require
enforcing transitivity constraints.  We summarize the results for the
transitivity constraints in our 100 buggy variants of \dlxcc{} in
terms of the minimum, the average, and the maximum of each
measurement.  We also show results for five synthetic benchmarks
consisting of $n \times n$ planar meshes $\meshn$, with $n$ ranging
from 4 to 8, where the mesh for $n=6$ is illustrated in Figure
\ref{mesh-figure}.  For all of the circuit benchmarks, the number of
cycles, although large, appears to be manageable.  Moreover, the
cycles have at most 4 edges.  The synthetic benchmarks, on the other
hand, demonstrate the exponential growth predicted as worst case
behavior.  The number of cycles grows quickly as the meshes grow
larger.  Furthermore, the cycles can be much longer, causing the
number of clauses to grow even more rapidly.

\subsection{Adding More Relational Variables}

Enumerating the transitivity constraints based on the variables
in $\tvarset$ runs the risk of generating a Boolean formula of
exponential size.  We can guarantee polynomial growth by considering a
larger set of relational variables.  In general, let $\tvarset'$ be
some set of relational variables such that $\tvarset \subseteq
\tvarset'$, and let $\ftrans(\tvarset')$ be the transitivity
constraint formula generated by enumerating the chord-free cycles in
the graph $G(\tvarset')$.
\begin{theorem}
If $\tvarset$ is the set of relational variables in $\fsat$ and
$\tvarset \subseteq \tvarset'$, then the formula $\fsat \land
\ftrans(\tvarset)$ is satisfiable if and only if 
$\fsat \land \ftrans(\tvarset')$ is satisfiable.
\label{superset-theorem}
\end{theorem}

We introduce a series of lemmas to prove this theorem.  For a
 propositional formula $F$ over a set of variables ${\cal A}$ and an
 assignment $\chi \colon {\cal A} \rightarrow \{0,1\}$, define the
 {\em valuation} of $F$ under $\asst$, denoted $\aeval{F}{\asst}$, to
 be the result of evaluating formula $F$ according to assignment
 $\asst$.  We first prove that we can extend any assignment over a set
 of relational variables to one over a superset of these variables
 yielding identical valuations for both transistivity constraint
 formulas.

\begin{lemma}
For any sets of relational variables $\tvarset_1$ and $\tvarset_2$ such that $\tvarset_1 \subseteq \tvarset_2$, and
for any assignment $\asst_1 \colon \tvarset_1 \rightarrow \{0, 1\}$, 
such that $\aeval{\ftrans(\tvarset_1)}{\asst_1} = 1$,
there is an assignment
$\asst_2 \colon \tvarset_2 \rightarrow \{0, 1\}$ such that
$\aeval{\ftrans(\tvarset_2)}{\asst_2} = 1$
\label{extend-assignment-lemma}
\end{lemma}

\begin{proof}
We consider the case where $\tvarset_2 = \tvarset_1 \cup \{e_{i,j}\}$.
 The general statement of the proposition then holds by induction on
 $|\tvarset_2| - |\tvarset_1|$.

Define assignment $\asst_2$ to be:
\begin{eqnarray*}
\asst_2(e) & = & 
\left \{
\begin{array}{ll}
\asst_1(e),  & e \not = e_{i,j} \\
1, & \mbox{Graph $G(\tvarset_1,\asst)$ has a path of 1-edges from node $i$ to node $j$.} \\
0, & \mbox{otherwise}
\end{array} \right . 
\end{eqnarray*}

We
consider two cases:
\begin{enumerate}
\item
If $\asst_2(e_{i,j}) = 0$, then any
cycle in $G(\tvarset_2,\asst_2)$ through $e_{i,j}$ must contain a
0-edge other than $e_{i,j}$.  Hence adding this edge does not
introduce any transitivity violations.
\item
If $\asst_2(e_{i,j}) =
1$, then there must be some path $P_1$ of 1-edges between nodes $i$
and $j$ in $G(\tvarset_1,\asst_1)$.  In order for the introduction of
1-edge $e_{i,j}$ to create a transitivity violation, there must also
be some path $P_2$ between nodes $i$ and $j$ in
$G(\tvarset_1,\asst_1)$ containing exactly one 0-edge.  But then we
could concatenate paths $P_1$ and $P_2$ to form a cycle in
$G(\tvarset_1,\asst_1)$ containing exactly one 0-edge, implying that
$\aeval{\ftrans(\tvarset_1)}{\asst_1} = 0$.  We conclude therefore
that adding 1-edge $e_{i,j}$ does not introduce any transitivity
violations.
\end{enumerate}
\end{proof}

\begin{lemma}
For $\tvarset_1 \subseteq \tvarset_2$ and
for any assignment $\asst_2 \colon \tvarset_2 \rightarrow \{0, 1\}$, 
such that $\aeval{\ftrans(\tvarset_2)}{\asst_2} = 1$,
we also have
$\aeval{\ftrans(\tvarset_1)}{\asst_2} = 1$
\label{shrink-assignment-lemma}
\end{lemma}

\begin{proof}
We note that any cycle in
$G(\tvarset_1,\asst_2)$ must be present in $G(\tvarset_2,\asst_2)$ and
have the same edge labeling.  Thus, if $G(\tvarset_2,\asst_2)$ has no
cycle with a single 0-edge, then neither does $G(\tvarset_1,\asst_2)$.
\end{proof}

We now return to the proof of Theorem \ref{superset-theorem}.

\begin{proof}
Suppose
that $\fsat \land \ftrans(\tvarset)$ is satisfiable, i.e., 
there is some assignment $\asst$ such that $\aeval{\fsat}{\asst} =
\aeval{\ftrans(\tvarset)}{\asst} = 1$.  Then by Lemma
\ref{extend-assignment-lemma} we can find an assignment $\asst'$ such
that $\aeval{\ftrans(\tvarset')}{\asst'} = 1$.  Furthermore, since the
construction of $\asst'$ by Lemma \ref{extend-assignment-lemma} preserves the values assigned
to all variables in $\tvarset$, and these are the only relational
variables occurring in $\fsat$, we can conclude that
$\aeval{\fsat}{\asst'} = 1$.  Therefore 
$\fsat \land \ftrans(\tvarset')$ is satisfiable.

Suppose on the other hand that $\fsat \land \ftrans(\tvarset')$
is satisfiable, i.e., there is some assignment
$\asst'$ such that $\aeval{\fsat}{\asst'} =
\aeval{\ftrans(\tvarset')}{\asst'} = 1$.  Then by Lemma
\ref{shrink-assignment-lemma}  we also have
$\aeval{\ftrans(\tvarset)}{\asst'} = 1$, and hence
$\fsat \land \ftrans(\tvarset)$ is satisfiable.
\end{proof}

Our goal then is to add as few relational variables as possible in
order to reduce the size of the transitivity formula.  We will
continue to use our path enumeration algorithm to generate the
transitivity formula.

\subsection{Dense Enumeration}

For the {\em dense} enumeration method,
let $\tvarset_N$ denote the set of variables $e_{i,j}$ for all values
of $i$ and $j$ such that $1 \leq i < j \leq N$.  Graph
$G(\tvarset_N)$ is a complete, undirected graph.  In this graph, any
cycle of length greater than three must have a chord.  Hence our algorithm will
enumerate transitivity constraints of the form $e_{[i,j]} \land
e_{[j,k]} \Rightarrow e_{[i,k]}$, for all distinct values of $i$, $j$,
and $k$.  The graph has $N(N-1)$ edges and $N(N-1)(N-2)/6$ chord-free
cycles, yielding a total of $N(N-1)(N-2)/2 = O(N^3)$ transitivity
constraints.

The  columns labeled ``Dense'' in Table \ref{chord-free-table}
show the complexity of this method for the benchmark circuits.  For
the smaller graphs \dlxc{}, \dlxct{}, $M_4$ and $M_5$, this method yields more
clauses than direct enumeration of the cycles in the original graph.
For the larger graphs, however, it yields fewer clauses.  The
advantage of the dense method is most evident for the mesh graphs,
where the cubic complexity is far superior to exponential.

\subsection{Sparse Enumeration}

We can improve on both of these methods by exploiting the sparse
structure of $G(\tvarset)$.  Like the dense method, we want to
introduce additional relational variables to give a set of variables
$\atvarset$ such that the resulting graph $G(\atvarset)$ becomes {\em
chordal} \cite{rose-jmaa70}.  That is, the graph has the property that
every cycle of length greater than three has a chord.

Chordal graphs have been studied extensively in the context of sparse
Gaussian elimination.  In fact, the problem of finding a minimum set
of additional variables to add to our set is identical to the problem
of finding an elimination ordering for Gaussian elimination that
minimizes the amount of fill-in.  Although this problem is NP-complete
\cite{yannakakis-sjadm81}, there are good heuristic solutions.  In
particular, our implementation proceeds as a series of elimination
steps.  On each step, we remove some vertex $i$ from the graph.  For
every pair of distinct, uneliminated vertices $j$ and $k$ such that
the graph contains edges $(i,j)$ and $(i,k)$, we add an edge $(j,k)$
if it does not already exist.  The original graph plus all of the added
edges then forms a chordal graph.  To choose which vertex to eliminate
on a given step, our implementation uses the simple heuristic of choosing the vertex
with minimum degree.  If more than one vertex has minimum degree, we
choose one that minimizes the number of new edges added.

The columns in Table \ref{chord-free-table} labeled ``Sparse''
show the effect of making the benchmark graphs chordal by this
method. Observe that this method gives superior results to either of
the other two methods.  In our implementation we have therefore
used the sparse method to generate all of the transitivity constraint formulas.

\section{SAT-Based Decision Procedures}

Most Boolean satisfiability (SAT) checkers take as input a
formula expressed in clausal form.  Each clause is a set of {\em
literals}, where a literal is either a variable or its complement.  A
clause denotes the disjunction of its literals.  The task of the
checker is to either find an assignment to the variables that
satisfies all of the clauses or to determine that no such assignment
exists.  We can solve the constrained satisfiability problem using a
conventional SAT checker by generating a set of clauses
$\transclauseset$ representing $\ftrans(\atvarset)$ and a set of
clauses $\formclauseset$ representing the formula $\fsat$.  We then
run the checker on the combined clause set $\formclauseset \cup
\transclauseset$ to find satisfying solutions to $\fsat \land
\ftrans(\atvarset)$.

\begin{table}
\begin{center}
\begin{tabular}{|l|l|cr|cr|r|}
\hline
\multicolumn{2}{|c|}{Circuit}  &
\multicolumn{2}{|c|}{$\formclauseset$} & 
\multicolumn{2}{|c|}{$\transclauseset \cup \formclauseset$} &
\multicolumn{1}{|c|}{Ratio} \\
\multicolumn{2}{|c|}{} &
\multicolumn{1}{|c}{Satisfiable?} & 
\multicolumn{1}{c|}{Secs.} &
\multicolumn{1}{|c}{Satisfiable?} & 
\multicolumn{1}{c|}{Secs.} & 
\multicolumn{1}{|c|}{} \\
\hline
\multicolumn{2}{|l|}{\dlxc}  & N & 3 & N & 4 & 1.4  \\
\multicolumn{2}{|l|}{\dlxct} & Y & 1 & N & 9 & N.A.  \\
\hline
\multicolumn{2}{|l|}{\dlxca}  & N & 176 & N & 1,275 & 7.2 \\
\multicolumn{2}{|l|}{\dlxcat} & Y & 3 & N & 896 & N.A. \\
\hline
\multicolumn{2}{|l|}{\dlxcc}  & N & 5,035 & N & 9,932 & 2.0  \\
\multicolumn{2}{|l|}{\dlxcct} & Y & 4 & N & 15,003 & N.A.  \\
\hline
Full  & min. & Y & 1 & Y & 1 & 0.2 \\
Buggy & avg. & Y & 125 & Y & 1,517 & 2.3  \\
\dlxcc & max.  & Y & 2,186 & Y & 43,817 & 69.4 \\
\hline
\end{tabular}
\end{center}
\mycaption{Performance of {\sc fgrasp} on Benchmark Circuits.}{Results are given both without and with transitivity
constraints.}
\label{sat-table}
\end{table}

In experimenting with a number of Boolean satisfiability checkers, we
have found that {\sc fgrasp} \cite{marques-tc99} has the best overall
performance.  The most recent version can be directed to periodically
restart the search using a randomly-generated variable assignment
\cite{marques-pcai99}.  This is the first SAT checker we have tested
that can complete all of our benchmarks.  All of our experiments were
conducted on a 336 MHz Sun UltraSPARC II with 1.2GB of primary memory.

As indicated by Table \ref{sat-table}, we ran {\sc fgrasp} on
clause sets $\formclauseset$ and $\transclauseset \cup \formclauseset$, i.e., 
both without and with transitivity constraints.
For benchmarks \dlxc{}, \dlxca{}, and \dlxcc{}, the formula $\fsat$ is
unsatisfiable.  As can be seen, including transitivity constraints increases the run time
significantly.  For benchmarks \dlxct{}, \dlxcat{}, and \dlxcct{}, the formula
$\fsat$ is satisfiable, but only because transitivity is not enforced.  When
we add the clauses for $\ftrans$, the formula becomes unsatisfiable.  For the buggy
circuits, the run times for $\formclauseset$ range from under 1 second to over 36 minutes.
The run times for $\transclauseset \cup \formclauseset$
range from less than one second to over 12 hours.  In some cases,
adding transitivity constraints actually decreased the CPU time (by as much as a factor of 5), but
in most cases the CPU time increased (by as much as a factor of 69).  On average (using the geometric mean) adding
transitivity constraints increased the CPU time by a factor of 2.3.
We therefore conclude that satisfiability checking with transitivity constraints is more difficult than conventional satisfiability
checking, but the added complexity is not overwhelming.

\section{OBDD-Based Decision Procedures}

A simple-minded approach to solving satisfiability with transitivity
constraints using OBDDs would be to generate separate OBDD
representations of $\ftrans$ and $\fsat$.  We could then
use the {\sc Apply} operation to generate an OBDD for
$\ftrans \land \fsat$, and then either find a satisfying
assignment or determine that the function is unsatisfiable.
We show that for some sets of relational variables $\tvarset$, the OBDD
representation of $\ftrans(\tvarset)$ can be too large to
represent and manipulate.
In our experiments, we use the CUDD OBDD
package with dynamic variable reordering by sifting.

\subsection{Lower Bound on the OBDD Representation of $\ftrans(\tvarset)$}

\begin{figure}[t]
\newcommand{\mvertex}[2]{\vertex \put(-.1,.1){\makebox(0,0)[br]{$v_{#1,#2}$}}}
\newcommand{\face}[2]{\clabel{$F_{#1,#2}$}}
\setlength{\unitlength}{.25in}
\begin{center}
\begin{picture}(12,12)
\put(1,1){\line(1,0){10}}
\put(1,3){\line(1,0){10}}
\put(1,5){\line(1,0){10}}
\put(1,7){\line(1,0){10}}
\put(1,9){\line(1,0){10}}
\put(1,11){\line(1,0){10}}
\put(1,1){\line(0,1){10}}
\put(3,1){\line(0,1){10}}
\put(5,1){\line(0,1){10}}
\put(7,1){\line(0,1){10}}
\put(9,1){\line(0,1){10}}
\put(11,1){\line(0,1){10}}
\put(1,11){\mvertex{1}{1}}
\put(3,11){\mvertex{1}{2}}
\put(5,11){\mvertex{1}{3}} 
\put(7,11){\mvertex{1}{4}} 
\put(9,11){\mvertex{1}{5}} 
\put(11,11){\mvertex{1}{6}} 
\put(1,9){\mvertex{2}{1}}
\put(3,9){\mvertex{2}{2}}
\put(5,9){\mvertex{2}{3}} 
\put(7,9){\mvertex{2}{4}} 
\put(9,9){\mvertex{2}{5}} 
\put(11,9){\mvertex{2}{6}} 
\put(1,7){\mvertex{3}{1}}
\put(3,7){\mvertex{3}{2}}
\put(5,7){\mvertex{3}{3}} 
\put(7,7){\mvertex{3}{4}} 
\put(9,7){\mvertex{3}{5}} 
\put(11,7){\mvertex{3}{6}} 
\put(1,5){\mvertex{4}{1}}
\put(3,5){\mvertex{4}{2}}
\put(5,5){\mvertex{4}{3}} 
\put(7,5){\mvertex{4}{4}} 
\put(9,5){\mvertex{4}{5}} 
\put(11,5){\mvertex{4}{6}} 
\put(1,3){\mvertex{5}{1}}
\put(3,3){\mvertex{5}{2}}
\put(5,3){\mvertex{5}{3}} 
\put(7,3){\mvertex{5}{4}} 
\put(9,3){\mvertex{5}{5}} 
\put(11,3){\mvertex{5}{6}} 
\put(1,1){\mvertex{6}{1}}
\put(3,1){\mvertex{6}{2}}
\put(5,1){\mvertex{6}{3}} 
\put(7,1){\mvertex{6}{4}} 
\put(9,1){\mvertex{6}{5}} 
\put(11,1){\mvertex{6}{6}} 
\put(2,10){\face{1}{1}}
\put(4,10){\face{1}{2}}
\put(6,10){\face{1}{3}}
\put(8,10){\face{1}{4}}
\put(10,10){\face{1}{5}}
\put(2,8){\face{2}{1}}
\put(4,8){\face{2}{2}}
\put(6,8){\face{2}{3}}
\put(8,8){\face{2}{4}}
\put(10,8){\face{2}{5}}
\put(2,6){\face{3}{1}}
\put(4,6){\face{3}{2}}
\put(6,6){\face{3}{3}}
\put(8,6){\face{3}{4}}
\put(10,6){\face{3}{5}}
\put(2,4){\face{4}{1}}
\put(4,4){\face{4}{2}}
\put(6,4){\face{4}{3}}
\put(8,4){\face{4}{4}}
\put(10,4){\face{4}{5}}
\put(2,2){\face{5}{1}}
\put(4,2){\face{5}{2}}
\put(6,2){\face{5}{3}}
\put(8,2){\face{5}{4}}
\put(10,2){\face{5}{5}}
\end{picture}
\end{center}
\mycaption{Mesh Graph $M_6$.}{}
\label{mesh-figure}
\end{figure}

We prove that for some sets $\tvarset$, the OBDD representation of
$\ftrans(\tvarset)$ may be of exponential size for all possible
variable orderings.  As mentioned earlier, the NP-completeness result
proved by Goel, {\em et al.}~\cite{goel-cav98} has implications for the
complexity of representing $\ftrans(\tvarset)$ as an OBDD{}.  They
showed that given an OBDD $G_{\mbox{sat}}$ representing formula
$\fsat$, the task of finding a satisfying assignment of $\fsat$ that
also satisfies the transitivity constraints in $\transset(\tvarset)$
is NP-complete in the size of $G_{\mbox{sat}}$.  By this, assuming
${\it P} \not = {\it NP}$, we can infer that the OBDD representation
of $\ftrans(\tvarset)$ may be of exponential size when using the
same variable ordering as is used in $G_{\mbox{sat}}$.  Our result extends this
lower bound to arbitrary variable orderings and is independent of the 
${\it P}$ vs.~${\it NP}$ problem.

Let $\meshn$ denote a planar mesh consisting of a square array of $n
\times n$ vertices.   For example, Figure
\ref{mesh-figure} shows the graph for $n = 6$.  Being a planar graph,
the edges partition the plane into {\em faces}.  As shown in Figure
\ref{mesh-figure} we label these $F_{i,j}$ for $1 \leq i,j \leq n-1$.
There are a
total of $(n-1)^2$ such faces.  One can see that the set of edges
forming the border of each face forms a chord-free cycle of $M_n$.  
As shown in Table \ref{chord-free-table}, many other cycles
are also chord-free, e.g., the perimeter of any rectangular region
having height and width greater than 1, but we will consider only the
cycles corresponding to single faces.  

Define $\meshvars$ to be a set of relational variables corresponding to the edges in $M_n$.
$\ftrans(\meshvars)$ is then an encoding of the transitivity constraints for these variables.

\begin{theorem}
Any OBDD representation of $\ftrans(\meshvars)$
must have $\Omega(2^{n/4})$ vertices.
\label{lower-bound-theorem}
\end{theorem}

\begin{figure}
\centerline{\psfig{figure=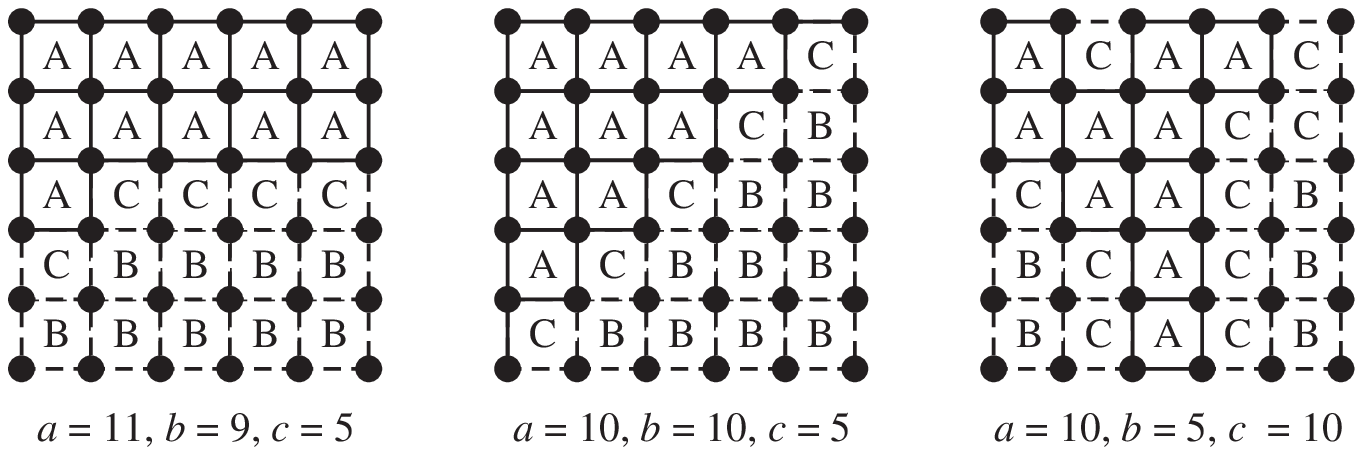,width=6.0in}}
\mycaption{Partitioning Edges into Sets $A$ (solid) and $B$ (dashed).}
{Each face can then be classified as type A (all solid), B (all dashed), or C (mixed).}
\label{grid-partition-fig}
\end{figure}

To prove this theorem, consider any ordering of the variables
representing the edges in $\meshn$.  Let $A$ denote those in the first
half of the ordering, and $B$ denote those in the second half.  We can
then classify each face according to the four edges forming its
border:
\begin{description}
\item[A:] All are in $A$.
\item[B:] All are in $B$.
\item[C:] Some are in $A$, while others are in $B$.  These are called
``split'' faces.
\end{description}
Observe that we cannot have a type A face adjacent to a type B face,
since their shared edge cannot be in both $A$ and $B$.  Therefore
there must be split faces separating any region of type A faces from any
region of type B faces.

For example, Figure \ref{grid-partition-fig} shows three possible
partitionings of the edges of $M_6$ and the resulting classification
of the faces.  If we let $a$, $b$, and $c$ denote the number of faces
of each respective type, we see that we always have $c \geq 5 = n-1$.
In particular, a minimum value for $c$ is achieved when the
partitioning of the edges corresponds to a partitioning of the graph
into a region of type A faces and a region of type B faces, each
having nearly equal size, with the split faces forming the boundary
between the two regions.

\begin{lemma}

For 
any partitioning of the edges of mesh graph $\meshn$ into equally-sized
sets $A$ and $B$, there must be at least $(n-3)/2$ split faces.
\label{bisection-lemma}
\end{lemma}

Note that this lower bound is somewhat weak---it seems clear that we
must have $c \geq n-1$.  However, this weaker bound will suffice to
prove an exponential lower bound on the OBDD size.

\begin{proof}
Our proof is an adaptation of a proof by Leighton
\cite[Theorem~1.21]{leighton-92} that $\meshn$ has a bisection
bandwidth of at least $n$.  That is, one would have to remove at least
$n$ edges to split the graph into two parts of equal size.

Observe that $\meshn$ has $n^2$ vertices and $2 n (n-1)$ edges.  These
edges are split so that $n(n-1)$ are in $A$ and $n(n-1)$ are in $B$.

Let $\dualmeshn$ denote the planar dual of $\meshn$.  That is, it
contains a vertex $u_{i,j}$ for each face $F_{i,j}$ of $\meshn$, and
edges between pairs of vertices such that the corresponding faces in
$\meshn$ have a common edge.  In fact, one can readily see that this
graph is isomorphic to $\mesh{n-1}$.

Partition the vertices of $\dualmeshn$ into sets $U_a$, $U_b$, and
$U_c$ according to the types of their corresponding faces.  Let $a$,
$b$, and $c$ denote the number of elements in each of these sets.
Each face of $\meshn$ has four bordering edges, and each edge is the
border of at most two faces.  Thus, as an upper bound on $a$, we must
have $4a \leq 2n(n-1)$, giving $a \leq n(n-1)/2$, and similarly for
$b$.  In addition, since a face of type A cannot be adjacent in
$\meshn$ to one of type B, no vertex in $U_a$ can be adjacent in
$\dualmeshn$ to one in $U_b$.

Consider the complete, directed, bipartite graph having as edges the
set $(U_a \times U_b) \cup (U_b \times U_a)$, i.e., a total of $2ab$
edges.  Given the bounds: $a + b = (n-1)^2 - c$, $a \leq n(n-1)/2$,
and $b \leq n(n-1)/2$, the minimum value of $2ab$ is achieved when
either $a = n(n-1)/2$ and $b = (n-1)^2 - (n-1)n/2 - c = (n-1)(n-2)/2 -
c$, or vice-versa, giving a lower bound:
\begin{eqnarray*}
2ab & \geq & 2[n(n-1)/2]\cdot[(n-1)(n-2)/2 - c] \\
& = & n(n-1)^2(n-2)/2\; - \;cn(n-1)
\end{eqnarray*}

We can embed this bipartite graph in $M^D_{n}$ by forming a path from
vertex $u_{i,j}$ to vertex $u_{i',j'}$, where either $u_{i,j} \in U_a$
and $u_{i',j'} \in U_b$, or vice-versa.  By convention, we will use
the path that first follows vertical edges to $u_{i',j}$ and then
follows horizontal edges to $u_{i',j'}$.  We must have at least one
vertex in $U_c$ along each such path, and therefore removing the
vertices in $U_c$ would cut all $2ab$ paths.

For each vertex $u_{i,j} \in U_c$, we can bound the total number of
paths passing through it by separately considering paths that enter
from the bottom, the top, the left, and the right.  For those entering
from the bottom, there are at most $n-i-1$ source vertices and
$i(n-1)$ destination vertices, giving at most $i(n-i-1)(n-1)$ paths.
This quantity is maximized for $i = (n-1)/2$, giving an upper bound of
$(n-1)^3/4$.  A similar argument shows that there are at most
$(n-1)^3/4$ paths entering from the top of any vertex.  For the
paths entering from the left, there are at most $(j-1)(n-1)$ source
vertices and $(n-j)$ destinations, giving at most $(j-1)(n-j)(n-1)$
paths.  This quantity is maximized when $j = (n-1)/2$, giving an upper
bound of $(n-1)^3/4$.  This bound also holds for those paths entering
from the right.  Thus, removing a single vertex would cut at most
$(n-1)^3$ paths.

Combining the lower bound on the number of paths $2ab$, the upper
bound on the number of paths cut by removing a single vertex, and the
fact that we are removing $c$ vertices, we have:
\begin{eqnarray*}
c (n-1)^3 &  \geq & n(n-1)^2(n-2)/2 - cn(n-1) \\
c (n-1)^3 + cn & \geq & n(n-1)(n-2)/2 \\
c (n^2 -n + 1) & \geq & n(n-1)(n-2)/2
\end{eqnarray*}
We can rewrite $n(n-1)(n-2)$ as $(n^2-n+1)(n-3) + n^2-2n+3$.  Observing that
$n^2-2n+3 > 0$ for all values of $n$, we have:
\begin{eqnarray*}
c (n^2 -n + 1) & \geq & (n^2-n+1)(n-3)/2 + (n^2-2n+3)/2\\
		& \geq & (n^2-n+1)(n-3)/2\\
c  & \geq & (n-3)/2
\end{eqnarray*}
\end{proof}

A set of faces is said to be {\em edge independent} when no two members of the
set share an edge.

\begin{lemma}
For
any partitioning of the edges of mesh graph $\meshn$ into equally-sized
sets $A$ and $B$, there must be an edge-independent set of  split faces
containing at least $(n-3)/4$ elements.
\end{lemma}

\begin{proof}
Classify the {\em parity} of face $F_{i,j}$ as ``even'' when $i + j$
is even, and as ``odd'' otherwise.  Observe that no two faces of the
same parity can have a common edge.  Divide the set of split faces
into two subsets: those with even parity and those with odd.  Both of
these subsets are edge independent, and one of them must have at least 1/2
of the elements of the set of all split faces.
\end{proof}

We can now complete the proof of
Theorem \ref{lower-bound-theorem}
\begin{proof}
Suppose
there is an edge-independent set of $k$ split faces.  For each split
face, choose one edge in $A$ and one edge in $B$ bordering that face.
For each value $\vec{y} \in \{0,1\}^k$, define assignment
$\alpha_{\vec{y}}$ (respectively, $\beta_{\vec{y}}$), to the variables
representing edges in $A$ (resp., $B$) as follows.  For an edge $e$
that is not part of any of the $k$ split faces, define
$\alpha_{\vec{y}}(e) = 0$ (resp., $\beta_{\vec{y}}(e) = 0$).  For an
edge $e$ that is part of a split face, but it was not one of the ones
chosen specially, let $\alpha_{\vec{y}} = 1$ (resp.,
$\beta_{\vec{y}}(e) = 1$).  For an edge $e$ that is the chosen
variable in face $i$, let $\alpha_{\vec{y}}(e) = y_i$ (resp.,
$\beta_{\vec{y}}(e) = y_i$).  This will give us an assignment
$\alpha_{\vec{y}} \cdot \beta_{\vec{y}}$ to all of the variables that
evaluates to 1.  That is, for each independent, split face $F_i$, we
will have two 1-edges when $y_i = 0$ and four 1-edges when $y_i = 1$.
All other cycles in the graph will have at least two 0-edges.

On the other hand, for any $\vec{y}, \vec{z} \in \{0,1\}^k$ such that
$\vec{y} \not = \vec{z}$ the
assignment
$\alpha_{\vec{y}}
\cdot
\beta_{\vec{z}}$ will cause an evaluation to 0, because for any face
$i$ where $y_i \not = z_i$,
all but one edge will be assigned value 1.
Thus, the set of
assignments $\{\alpha_{\vec{y}}|\vec{y} \in \{0,1\}^k\}$ forms an OBDD
fooling set, as defined in \cite{bryant-ieeetc91}, implying that the OBDD
must have at least
$2^k \geq 2^{(n-3)/4} = \Omega(2^{n/4})$ vertices.
\end{proof}

We have seen that adding relational variables can reduce the number of
cycles and therefore simplify the transitivity constraint formula.
This raises the question of how adding relational variables affects
the BDD representation of the transitivity constraints.
Unfortunately, the exponential lower bound still holds.

\begin{corollary}
For any set of relational variables $\tvarset$ such that $\meshvars
\subseteq \tvarset$, any OBDD representation of $\ftrans(\tvarset)$ must
contain $\Omega(2^{n/8})$ vertices.
\label{lower-bound-corollary}
\end{corollary}

The extra edges in $\tvarset$ introduce complications, because they
create cycles containing edges from different faces.  As a result, our
lower bound is weaker.

Define a set of faces as {\em vertex independent} if no two members
share a vertex.
\begin{lemma}
For
any partitioning of the edges of mesh graph $\meshn$ into equal-sized
sets $A$ and $B$, there must be a vertex-independent set of  split faces
containing at least $(n-3)/8$ elements.
\label{vertex-split-lemma}
\end{lemma}

\begin{proof}
Partition the set of split faces into four sets: EE, EO, OE, and OO,
where face $F_{i,j}$ is assigned to a set according to the values of
$i$ and $j$:
\begin{enumerate}
\item[EE:] Both $i$ and $j$ are even.
\item[EO:] $i$ is even and $j$ is odd.
\item[OE:] $i$ is odd and $j$ is even.
\item[OO:] Both $i$ and $j$ are odd.
\end{enumerate}
Each of these sets is vertex independent.  At least one of the sets
must contain at least $1/4$ of the elements.  Since there are at
least $(n-3)/2$ split faces, one of the sets must contain at
least $(n-3)/8$ vertex-independent split faces.
\end{proof}

We can now prove Corollary \ref{lower-bound-corollary}.

\begin{proof}
For any ordering of the variables in $\tvarset$, partition them into
two sets $A$ and $B$ such that those in $A$ come before those in $B$,
and such the number of variables that are in $\meshvars$
are equally split between $A$ and $B$.  Suppose there
is a vertex-independent set of $k$ split faces.  For each value
$\vec{y} \in \{0,1\}^k$, we define assignments $\alpha_{\vec{y}}$ to
the variables in $A$ and $\beta_{\vec{y}}$ to the variables in $B$.
These assignments are defined as they are in the proof of Theorem
\ref{lower-bound-theorem} with the addition that each variable
$e_{i,j}$ in $\tvarset - \meshvars$ 
is assigned value 0.  
Consider the set of assignments $\alpha_{\vec{y}}\cdot\beta_{\vec{z}}$
for all values $\vec{y},\vec{z} \in \{0,1\}^k$.
The only cycles in $G(\tvarset,\alpha_{\vec{y}}\cdot\beta_{\vec{z}})$
that can have less than two 0-edges will be those corresponding to the
perimeters of split faces.  
As in the proof of Theorem \ref{lower-bound-theorem}, the set 
$\{\alpha_{\vec{y}}|\vec{y} \in \{0,1\}^k\}$ forms an OBDD
fooling set, as defined in \cite{bryant-ieeetc91}, implying that the OBDD
must have at least
$2^k \geq 2^{(n-3)/8} = \Omega(2^{n/8})$ vertices.
\end{proof}

Our lower bounds are fairly weak, but this is more a reflection of the
difficulty of proving lower bounds.  We have found in practice that the OBDD
representations of the transitivity constraint functions arising from
benchmarks tend to be large relative to those encountered during the
evaluation of $\fsat$.  For example, although the OBDD representation
of $\ftrans(\atvarset)$ for benchmark \dlxct{}
is just 2,692 nodes (a
function over 42 variables), we have been unable to construct the OBDD
representations of this function for either \dlxcat{} (178 variables)
or \dlxcct{} (193 variables) despite running for over 24 hours.

\subsection{Enumerating and Eliminating Violations}

Goel, {\em et al.} \cite{goel-cav98} proposed a method that generates
implicants (cubes) of the function $\fsat$ from its OBDD
representation.  Each implicant is examined and discarded if it
violates a transitivity constraint.  In our experiments, we have found
this approach works well for the normal, correctly-designed pipelines
(i.e., circuits \dlxc{}, \dlxca{}, and \dlxcc{}) since the formula
$\fsat$ is unsatisfiable and hence has no implicants.  For all 100 of
our buggy circuits, the first implicant generated contained no
transitivity violation and hence was a valid counterexample.

For circuits that do require enforcing transitivity constraints, we
have found this approach impractical.  For example, in verifying
\dlxct{} by this means, we generated 253,216 implicants, requiring a
total of 35 seconds of CPU time (vs.~0.2 seconds for \dlxc{}).  For
benchmarks \dlxcat{} and \dlxcct{}, our program ran for over 24 hours
without having generated all of the implicants.  By contrast, circuits
\dlxca{} and \dlxcc{} can be verified in 11 and 29 seconds,
respectively.  Our implementation could be improved by making sure
that we generate only implicants that are irredundant and prime.  In
general, however, we believe that a verifier that generates individual
implicants will not be very robust.  The complex control
logic for a pipeline can lead to formulas $\fsat$ containing very
large numbers of implicants, even when transitivity plays only a minor
role in the correctness of the design.

\subsection{Enforcing a Reduced Set of Transitivity Constraints}

\begin{table}
\begin{center}
{\small
\begin{tabular}{|l|l|r|rrr|rrr|rrr|}
\hline
\multicolumn{2}{|c|}{Circuit}  &
\multicolumn{1}{|c|}{Verts.} &
\multicolumn{3}{|c|}{Direct} &
\multicolumn{3}{|c|}{Dense} & \multicolumn{3}{|c|}{Sparse} \\
\multicolumn{2}{|c|}{} && \multicolumn{1}{|c}{Edges} & 
\multicolumn{1}{c}{Cycles} & 
\multicolumn{1}{c|}{Clauses} &
\multicolumn{1}{|c}{Edges} & 
\multicolumn{1}{c}{Cycles} & 
\multicolumn{1}{c|}{Clauses} & 
\multicolumn{1}{|c}{Edges} & 
\multicolumn{1}{c}{Cycles} & 
\multicolumn{1}{c|}{Clauses} \\
\hline
\multicolumn{2}{|l|}{\dlxct} &  9 & 18 & 14 & 45 & 36 & 84 & 252 & 20 & 19 & 57 \\
\multicolumn{2}{|l|}{\dlxcat}& 17 & 44 & 101 & 395 & 136 & 680 & 2,040 & 49 & 57 & 171 \\
\multicolumn{2}{|l|}{\dlxcct}& 17 & 46 & 108 & 417 & 136 & 680 & 2,040 & 52 & 66 & 198 \\ 
\hline
Reduced  & min. & 3 & 2 & 0 & 0 & 3 & 1 & 3 & 2 & 0 & 0 \\
Buggy & avg. & 12 & 17 & 19 & 75 & 73 & 303 & 910 & 21 & 14 & 42 \\
\dlxcc{} & max. & 19 & 52 & 378 & 1,512 & 171 & 969 & 2,907 & 68 & 140 & 420 \\
\hline
\end{tabular}
} 
\end{center}
\mycaption{Graphs for Reduced Transitivity Constraints.}{Results are given for the three different methods
of encoding transitivity constraints based on the variables in the true support of $\fsat$.}
\label{reduced-table}
\end{table}

One advantage of OBDDs over other representations of Boolean functions
is that we can readily determine the {\em true support} of the function,
i.e., the set of variables on which the function depends.  This leads
to a strategy of computing an OBDD representation of $\fsat$ and
intersecting its support with $\tvarset$ to give a set $\rtvarset$ of
relational variables that could potentially lead to transitivity
violations.  We then augment these variables to make the graph
chordal, yielding a set of variables $\artvarset$ and
generate an OBDD representation of $\ftrans(\artvarset)$.  We
compute $\fsat \land \ftrans(\artvarset)$ and, if it is
satisfiable, generate a counterexample.

Table \ref{reduced-table} shows the complexity of the graphs generated
by this method for our benchmark circuits.  Comparing these with the
full graphs shown in Table \ref{chord-free-table}, we see that we
typically reduce the number of relational vertices (i.e., edges) by a
factor of 3 for the benchmarks modified to require transitivity and by
an even greater factor for the buggy circuit benchmarks.  The
resulting graphs are also very sparse.  For example, we can see that
both the direct and sparse methods of encoding transitivity
constraints greatly outperform the dense method.

\begin{table}
\begin{center}
\begin{tabular}{|l|l|rrr|r|}
\hline
\multicolumn{2}{|c|}{Circuit} & 
\multicolumn{3}{|c|}{OBDD Nodes} &
\multicolumn{1}{|c|}{CPU} \\
\multicolumn{2}{|c|}{} & 
\multicolumn{1}{|c}{$\fsat$} &
\multicolumn{1}{c}{$\ftrans(\artvarset)$} &
\multicolumn{1}{c|}{$\fsat \land \ftrans(\artvarset)$} &
Secs.\\
\hline
\multicolumn{2}{|l|}{\dlxc}	 &	1 &	1  & 1 &	0.2 \\
\multicolumn{2}{|l|}{\dlxct} &	530 &	344  &	1 &2 \\
\hline
\multicolumn{2}{|l|}{\dlxca} &	1 &	1  &	1 & 11 \\
\multicolumn{2}{|l|}{\dlxcat}& 22,491 & 10,656 &	1 &109 \\
\hline
\multicolumn{2}{|l|}{\dlxcc}	&	1 &	1  &	1 &29 \\
\multicolumn{2}{|l|}{\dlxcct}& 17,079 & 7,168 &	1 &441 \\

\hline
Reduced & min. & 20	& 1	& 20  & 7\\
 Buggy  & avg. & 3,173 & 1,483 &  25,057 & 107 \\
 \dlxcc{} & max. & 15,784 & 93,937  & 438,870 & 2,466\\
\hline
\end{tabular}
\end{center}
\mycaption{OBDD-based Verification.}{Transitivity constraints were generated for a reduced set of variables $\protect{\rtvarset}$.}
\label{bdd-table}
\end{table}

Table \ref{bdd-table} shows the complexity of applying the OBDD-based method to all
of our benchmarks.  The original circuits \dlxc{}, \dlxca{}, and
\dlxcc{} yielded formulas $\fsat$ that were unsatisfiable, and hence
no transitivity constraints were required.  The 3 modified circuits
\dlxct{}, \dlxcat{}, and \dlxcct{} are more interesting. 
The reduction in the number of relational variables
 makes it feasible to generate an OBDD representation of
the transitivity constraints.  Compared to benchmarks \dlxc{},
\dlxca{}, and \dlxcc{}, we see there is a significant, although
tolerable, increase in the computational requirement to verify the
modified circuits.  This can be attributed to both the more complex
control logic and to the need to apply the transitivity constraints.

For the 100 buggy variants of \dlxcc, 
$\fsat$ depends on up to 52 relational variables, with an
average of 17.  This yielded OBDDs
for $\ftrans(\artvarset)$ ranging up to 93,937 nodes, with an
average of 1,483.  The OBDDs for
$\fsat \land \ftrans(\artvarset)$ ranged up to 438,870
nodes (average 25,057), showing that adding transitivity constraints
does significantly increase the complexity of the OBDD representation.
However, this is just one OBDD at the end of a sequence of OBDD
operations.  In the worst case, imposing transitivity
constraints increased the total CPU time by a factor of 2, but on
average it only increased by 2\%.  The memory required to generate
$\fsat$ ranged from 9.8 to 50.9 MB (average 15.5), but even in the
worst case the total memory requirement increased by only 2\%.

\section{Conclusion}

By formulating a graphical interpretation of the relational variables,
we have shown that we can generate a set of clauses expressing the
transitivity constraints that exploits the sparse structure of the
relation.  Adding relational variables to make the graph chordal
eliminates the theoretical possibility of there being an exponential
number of clauses and also works well in practice.   A conventional SAT checker can
then solve constrained satisfiability problems, although the run times increase
significantly compared to unconstrained satisfiability.
Our best results were obtained
using OBDDs.  By considering only the relational variables in the
true support of $\fsat$, we can enforce transitivity constraints with only
a small increase in CPU time.

\end{document}